\documentclass[10pt,letterpaper]{article}

\usepackage{amsmath,amssymb}
\usepackage{changepage}
\usepackage[utf8x]{inputenc}
\usepackage{textcomp,marvosym}
\usepackage{cite}
\usepackage{nameref,hyperref}
\usepackage[right]{lineno}
\usepackage{microtype}
\DisableLigatures[f]{encoding = *, family = * }
\usepackage[table]{xcolor}
\usepackage{array}

\newcolumntype{+}{!{\vrule width 2pt}}

\newlength\savedwidth

\usepackage[aboveskip=1pt,labelfont=bf,labelsep=period,justification=raggedright,singlelinecheck=off]{caption}

\makeatletter
\renewcommand{\@biblabel}[1]{\quad#1.}
\makeatother

\date{}

\usepackage{booktabs}
\usepackage{placeins} 
\usepackage[font=small,labelfont=bf]{caption} 
\usepackage{floatpag} 
\usepackage{adjustbox}

\begin{document}
\vspace*{0.2in}

\begin{flushleft}
{\Large
\textbf\newline{Faster indicators of dengue fever case counts using \textit{Google} and \textit{Twitter}}}
\newline
\\
Giovanni Mizzi\textsuperscript{1,2*},
Tobias Preis\textsuperscript{2,3},
Leonardo Soares Bastos\textsuperscript{4},
Marcelo Ferreira da Costa Gomes\textsuperscript{4},
Claudia Torres Code\c{c}o\textsuperscript{4},
Helen Susannah Moat\textsuperscript{2,3}
\\
\bigskip
\textbf{1} Centre for Complexity Science, University of Warwick, Gibbet Hill Road, CV4 7AL, Coventry, United Kingdom
\\
\textbf{2} Data Science Lab, Behavioural Science, Warwick Business School, University of Warwick, Scarman Road, CV4 7AL, Coventry, United Kingdom
\\
\textbf{3} The Alan Turing Institute, British Library, 96 Euston Road, London, NW1 2DB, United Kingdom
\\
\textbf{4} Scientific Computing Program, Oswaldo Cruz Foundation, Av. Brazil, 4365 - Manguinhos, Rio de Janeiro, Brazil
\\
\bigskip
* giovanni.mizzi@gmail.com
\end{flushleft}

\section*{Abstract}

Dengue is a major threat to public health in Brazil, the world’s sixth biggest country by population, with over 1.5 million cases recorded in 2019 alone. Official data on dengue case counts is delivered incrementally and, for many reasons, often subject to delays of weeks. In contrast, data on dengue-related \emph{Google} searches and \emph{Twitter} messages is available in full with no delay. Here, we describe a model which uses online data to deliver improved weekly estimates of dengue incidence in Rio de Janeiro. We address a key shortcoming of previous online data disease surveillance models by explicitly accounting for the incremental delivery of case count data, to ensure that our approach can be used in practice. We also draw on data from \emph{Google Trends} and \emph{Twitter} in tandem, and demonstrate that this leads to slightly better estimates than a model using only one of these data streams alone. Our results provide evidence that online data can be used to improve both the accuracy and precision of rapid estimates of disease incidence, even where the underlying case count data is subject to long and varied delays.

\section*{Introduction}

Dengue is the most common mosquito-borne disease worldwide, with 50 to 100 million cases reported each year \cite{StanawayEtAl2016} and almost 4 billion people at risk \cite{BradyEtAl2012}. Typical symptoms of dengue include high fever, rashes, muscle aches and joint pain. A small proportion of patients develop a severe dengue infection, known as dengue haemorrhagic fever (DHF), which can involve massive bleeding and lead to death \cite{who2009}. The annual global number of dengue infections continues to grow, having already risen by a factor of 30 over the last 50 years \cite{WHO2012}. Unfortunately, there is currently no antiviral treatment to reduce severe illness \cite{Endy2014}, nor an effective vaccine.

Dengue has been endemic in Brazil since 1986, with all four serotypes circulating since 2012. Large epidemics occur every three to five years, causing disruption in the health system. With the high incidence rate of the disease, dengue is not only life-threatening, but also a serious burden on the Brazilian economy. To help mitigate dengue outbreaks, policymakers would greatly benefit from accurate, rapidly available information on the current number of cases of the disease. 

In reality, official data on the number of dengue cases is often delayed. In Brazil, some of these delays are caused by a lack of dedicated staff to complete the notification paperwork, as well as poor infrastructure in healthcare settings. While Brazil has an online reporting system, healthcare centres often do not have a good internet connection. In such situations, notification of each dengue case is often recorded on a paper sheet, which is then filed locally and sent to the municipal or state health secretariat for online submission. Delays are worsened further when surveillance teams are involved in other emergencies. 

In recent years, researchers have started to look at alternative sources of data which may provide rapid indicators of disease case counts. Instead of forecasting the incidence of the disease, the goal here is to ``nowcast'' the current number of cases, before the delayed official data is released. Previous work has investigated whether rapidly available data on people searching for a disease on \emph{Google} or discussing the disease on \emph{Twitter} could provide rapid insights into the incidence of a disease. For example, data on \emph{Google} searches has been shown to improve nowcasts of influenza case counts, in comparison to a model that makes estimates using official data alone \cite{Ginsberg2009,Preis2014,Yang2015,Lampos2015}. For dengue, relationships have been found between case counts and the use of online services such as \emph{Google} and \emph{Twitter} \cite{Gomide2011,Chan2011,Souza2015,Marques2017,Yang2017}, complementing other work that has sought to use rapidly available weather data \cite{Luz2008,Hii2012,Ramadona2016}.

However, delayed official data on the number of cases of a disease is often made available incrementally. For example, in Rio de Janeiro, around 25\% of dengue cases are entered into the system after a week, and less than 50\% after two weeks. Previous analyses of the value of online data in nowcasting dengue have not taken this incremental delivery into account, modelling official dengue data releases as lagged, full releases by working at a lower temporal granularity such as months \cite{Yang2017}.

Here, we seek to investigate whether online data can help improve weekly dengue case count nowcasts in a more realistic scenario where the official data is released incrementally. To do this, we build on a time series analysis framework for generating nowcasts of current disease case counts using historic, incrementally released case count data, introduced by Bastos and colleagues \cite{Bastos}. In addition, in contrast to previous approaches, we draw on data from \emph{Google Trends} and \emph{Twitter} in tandem, to investigate whether combining these two data sources can lead to better estimates than only using one at a time. We examine whether online data can improve both the accuracy and the precision of nowcasting estimates. The model we present is designed to be used in practice in the surveillance system \emph{InfoDengue}, which serves to detect dengue outbreaks in hundreds of Brazilian cities based on weekly official data \cite{codeco2016}.

\section*{Materials and Methods}
In this section, we detail the data sources used and the models analysed in the present study. The models that we consider all seek to deliver weekly estimates of dengue case counts in Rio de Janeiro. We carry out our analysis on the basis of epidemiological weeks, which are defined as starting on a Sunday.

\subsection*{Data sources}
Here we describe the three main sources of data used in this study.

\begin{description}
\item[Official data on dengue cases.] This is a list of suspected dengue cases for the city of Rio de Janeiro during the period from 1st January 2012 to 23rd July 2016.  Each case has a \emph{date of notification} and a \emph{date of system entry}. The \emph{date of notification} is the date on which the patient visits the doctor and dengue is diagnosed. The \emph{date of system entry} is the date on which the information about this case is inserted into the official database and becomes available for analysis, for example in nowcasting models such as those described here.
Note that suspected dengue cases later confirmed to be a disease other than dengue are removed from the list. The data was obtained from the Health Secretariat of Rio de Janeiro, via the \emph{InfoDengue} project.

\item[Google Trends.] Data on search behaviour was obtained via the \emph{Google Extended Trends API for Health}. We obtained daily data for the whole period of analysis from 1st January 2012 to 23rd July 2016. In order to identify searches relating to the topic of \emph{dengue}, we searched for the topic using \emph{Wikidata}\footnote{\href{https://www.wikidata.org/}{https://www.wikidata.org/}}, and then used the identified topic's Freebase identifier to query the \emph{Google Extended Trends API for Health}. For the topic of \emph{dengue fever} (referred to as \emph{dengue} from now on), the Freebase ID is \emph{/m/09wsg}. We chose the topic \emph{dengue fever} rather than \emph{dengue virus} as search volume for the latter was much lower. In Brazil, the finest geographical resolution for data retrieved from the \textit{Google Extended Trends API for Health} is state level. We therefore requested data on searches made in the state of Rio de Janeiro only. The data then returned by the API represents the probability of a few consecutive searches relating to dengue, including typos and indirect descriptions of the disease, within the state of Rio de Janeiro on each day in the period of analysis. 

Since 2015, the \textit{Zika} arbovirus has presented an additional risk in Rio de Janeiro, with considerable media coverage in some years. This disease is spread by the same mosquito as dengue, and also shares some symptoms. The same is true of a further arbovirus, \textit{chikungunya}, which has also been present in Rio de Janeiro since 2015, although with lower case counts. To allow us to investigate whether data on \textit{Google} searches relating to these two arboviruses might act as an additional potential signal for dengue incidence, we also retrieve searches relating to the topics of \emph{Zika virus} (referred to as \emph{Zika} from now on, Freebase ID \emph{/m/080m\_5j}; chosen instead of \emph{Zika fever} due to higher search volume) and \emph{chikungunya} (Freebase ID \emph{/m/01\_\_7l}). 

\item[Twitter.] We also analyse data on the volume of tweets relating to dengue that were posted to \emph{Twitter} during each week between 1st January 2012 and 23rd July 2016, for which the user location was determined to be in Rio de Janeiro city. Location was inferred on the basis of the user location specified in the \emph{Twitter} user's user information, as described in more detail by Gomide et al. \cite{Gomide2011}. The data reflects the volume of tweets that meet both the criteria of containing the word `dengue' and expressing personal experience of dengue (e.g., in English, ``You know I have had dengue?'') \cite{Gomide2011}. This dataset was made available to us by the \emph{Observatorio da Dengue}\footnote{\href{http://www.observatorio.inweb.org.br/dengue/}{http://www.observatorio.inweb.org.br/dengue/}} via the \emph{InfoDengue} project. 
\end{description}

\begin{figure}[!h]
\thisfloatpagestyle{empty}
\centering 
\includegraphics[width = \linewidth]{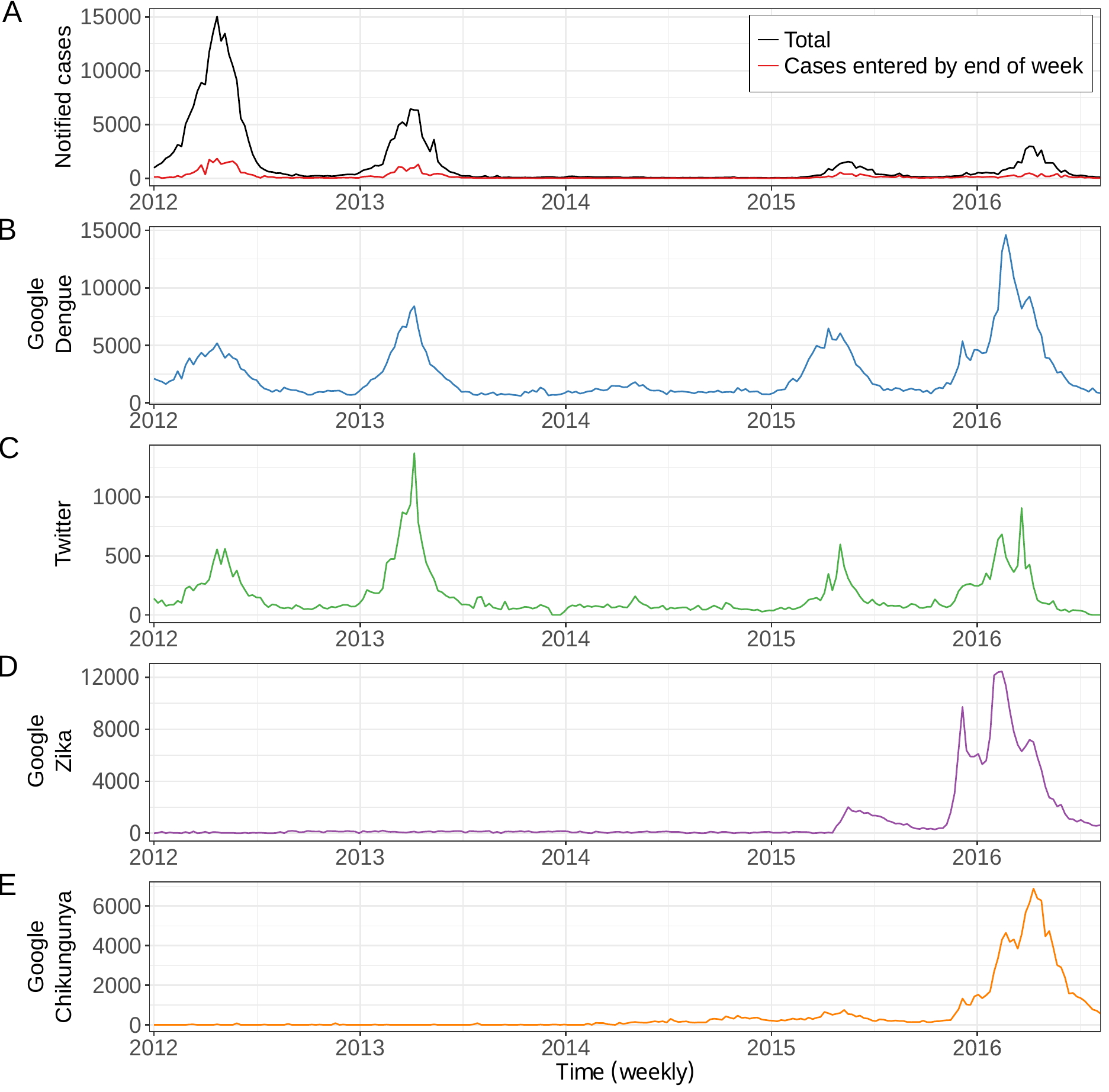}
\caption{{\bf Dengue case count data compared to data from \textit{Google} and \textit{Twitter}.} 
(A) In black, we depict official data on the total number of dengue cases recorded in official data for each week in Rio de Janeiro, from January 2012 until July 2016. The city frequently experiences dengue seasons during which thousands of people are infected. In red, we depict the total number of dengue cases known to the authorities by the end of each week. It is clear that only a very small fraction of dengue cases are entered into the database by the end of each week (see Fig. \ref{fig:Fig2} for further details). (B) We investigate whether rapidly available data on \textit{Google} searches relating to dengue can help improve our understanding of the number of dengue cases in the previous week. It can be seen that peaks in dengue related searches occur at roughly the same time as peaks in dengue cases. However, we note that the size of the peak in searches often does not directly correspond to the size of the peak in dengue cases. (C) We also examine the relationship between dengue case counts and the number of tweets in the city of Rio de Janeiro that express personal experience of dengue. Again, we see that peaks in tweets occur at roughly the same time as peaks in cases, but the relative size of the peaks does not always correspond. (D) Since 2015, the \textit{Zika} arbovirus has presented an additional risk in Rio de Janeiro, with considerable media coverage in some years. This disease is spread by the same mosquito as dengue, and also shares some symptoms. We therefore also investigate whether data on \textit{Google} searches relating to Zika might act as an additional potential signal for dengue incidence. (E) For similar reasons, we also consider data on \textit{Google} searches relating to the arbovirus \textit{chikungunya}. In Brazil, \textit{Google} data is made available via the \textit{Google Extended Trends API for Health} at state level and therefore relates to searches in the state of Rio de Janeiro.}
\label{fig:Fig1}
\end{figure}

\begin{figure}[!h]
\thisfloatpagestyle{empty}
\centering 
\adjustbox{valign=t}{\begin{minipage}[c]{0.55\linewidth}
    \includegraphics[width = \linewidth]{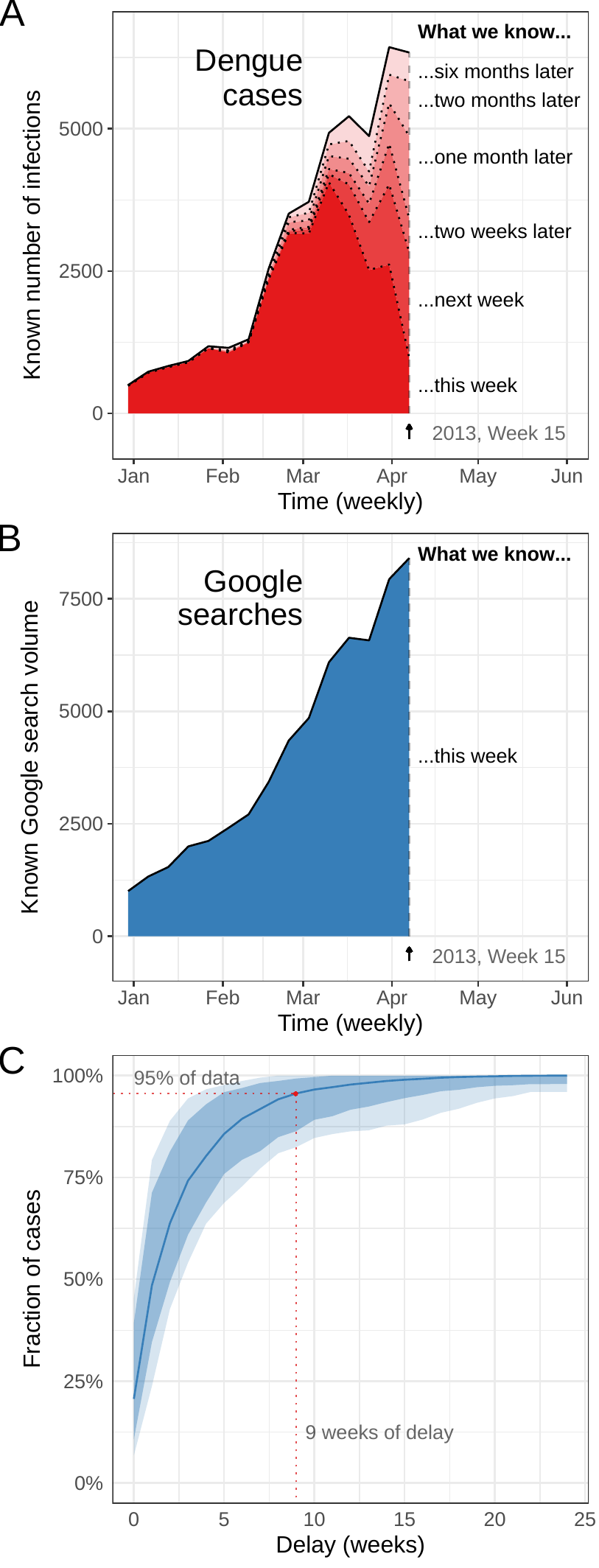}
\end{minipage}}\hfill
\adjustbox{valign=t}{\begin{minipage}[c]{0.42\linewidth}
    \caption{{\bf Delays in official data on dengue case counts.}
(A) We examine the true nature of the delays in the availability of official data on cases of dengue in Rio de Janeiro. We consider data from the 15\textsuperscript{th} epidemiological week of 2013 as an example. It can be seen that only a very small fraction of dengue cases have been entered into the surveillance system by the end of the week. Indeed, data relating to this week continues to arrive over a period of six months. Furthermore, by the end of the 15\textsuperscript{th} epidemiological week of 2013, data on dengue cases in the previous weeks is also severely incomplete. This creates problems for autoregressive methods that seek to use complete knowledge about previous weeks to compensate for delays in the arrival of data relating to the current week. (B) In contrast to official data on dengue cases, data on \textit{Google} searches in the 15\textsuperscript{th} epidemiological week of 2013 is available in full by the end of the week. The same applies to data on tweets posted on \textit{Twitter}. This opens up possibilities to use data on \textit{Google} searches and tweets relating to dengue to improve estimates of the number of dengue cases in a given week. (C) We further examine the rate with which dengue cases for a given week are added into the system. Here, we depict the empirical distribution of the delays in dengue case count entry over the whole time series. The blue line depicts the mean fraction of cases entered into the system after a given delay. The dark shading indicates 80\% of the empirical distribution of the fraction of cases notified after a given delay, and the light shading 95\% of the empirical distribution. It can be seen that there is a mean delay of 9 weeks before 95\% of dengue cases for a given week are entered into the system.}
    \label{fig:Fig2}
\end{minipage}}
\end{figure}

We depict all the time series described above in Fig. \ref{fig:Fig1}. It is possible to see that there is a correlation between the number of cases of dengue notified to doctors in a given week (Fig. \ref{fig:Fig1}A, black) and both the volume of \emph{Google} searches (Fig. \ref{fig:Fig1}B; Kendall's $\tau = 0.506$, $N=238$, $p<0.001$) and tweets (Fig. \ref{fig:Fig1}C; Kendall's $\tau = 0.557$, $N=238$, $p<0.001$) relating to the topic of dengue. Whereas data on \emph{Google} searches and tweets is available almost immediately, only a very small fraction of dengue cases are entered into the surveillance system and therefore known to policymakers and analysts in the same week in which the patient visits the doctor (Fig. \ref{fig:Fig1}A, red). Indeed, there is a mean delay of 9 weeks before 95\% of the cases notified to doctors in a given week are entered into the system (Fig. \ref{fig:Fig2}). This means that in any given week, the official data on dengue cases in previous weeks is also notably incomplete. This presents clear obstacles for autoregressive models that seek to infer the number of cases in a given week by drawing on complete knowledge about previous weeks. It can also be seen that the number of cases entered into the system in the same week in which the patient visited the doctor cannot simply be multiplied by a given constant in order to determine the total number of cases notified to doctors in that week (Fig. \ref{fig:Fig1}A).

\FloatBarrier

\FloatBarrier

For the reasons outlined when introducing the \emph{Google Trends} data above, we also examine the volume of \emph{Google} searches for the topics of \emph{Zika} (Fig. \ref{fig:Fig1}D) and \emph{chikungunya} (Fig. \ref{fig:Fig1}E). We find a correlation between the number of dengue cases notified to doctors in a given week and \emph{Google} searches for both \emph{Zika} and \emph{chikungunya} in the same week, both when considering the whole period of analysis (\emph{Zika} searches: Kendall's $\tau = 0.127$, $N=238$, $p<0.01$; \emph{chikungunya} searches: Kendall's $\tau = -0.09$, $N=238$, $p<0.05$) and the period beginning in the 1\textsuperscript{st} epidemiological week in 2015, the year in which Zika and chikungunya became present in Rio de Janeiro (\emph{Zika} searches: Kendall's $\tau = 0.499$, $N=81$, $p<0.001$; \emph{chikungunya} searches: Kendall's $\tau = 0.526$, $N=81$, $p<0.001$).

\subsection*{Models}
We investigate whether rapidly available data on \emph{Google} searches and tweets relating to dengue or other arboviruses present in Rio de Janeiro can enhance weekly estimates of the number of cases of dengue in Rio reported to doctors in the previous week. Importantly, we carry out these investigations while taking into account the incremental delivery of dengue case count data described in the previous section. We therefore compare the following seven models.

\begin{description}
\item[Baseline.] We first consider a model developed by Bastos \emph{et al.} \cite{Bastos} that aims to infer the number of cases of dengue in the previous week using the delayed dengue case count alone. In simple terms, the model aims to estimate the number of cases of dengue that will be reported for each week with a given number of weeks delay. The approach therefore explicitly models the gradual delivery of information relating to dengue cases in a given week over the following weeks.

Formally, let $n_{t,\tau}$ be the number of cases that occurred in week $t$ and were reported in week $t+\tau$, thus with delay $\tau$. 
We assume that $n_{t,\tau}$ follows a negative binomial distribution
\begin{equation*}
n_{t,\tau} \sim \mathcal{NB}(\lambda_{t,\tau},\phi), \quad t=0,1,2,\ldots \quad \tau=0,1,2,\ldots
\label{eqn:nttau}
\end{equation*}
which has the following form
\begin{equation*}
P(n_{t,\tau} = k) = \binom{\lambda_{t,\tau}+k-1}{k}(1-\phi)^{\lambda_{t,\tau}}\phi^k, \quad k=0,1,2,\ldots
\label{eqn:negbin}
\end{equation*}
where the mean $\lambda_{t,\tau}$ is given by
\begin{equation*}
\log{(\lambda_{t,\tau})} = \mu + \alpha_t + \beta_{\tau},
\label{eqn:lambdattau}
\end{equation*}
$\mu$ is a constant and $\alpha_t$ and $\beta_{\tau}$ are random effects with an autoregressive structure
\begin{equation*}
\begin{split}
\alpha_t & \sim \alpha_{t-1} + \mathcal{N}(0, \eta_{\alpha}),\\
\beta_{\tau} & \sim \beta_{\tau-1} + \mathcal{N}(0, \eta_{\beta}).
\end{split}
\label{eqn:atbtau}
\end{equation*}

Parameters are fit using the Integrated Nested Laplace Approximation (INLA) method \cite{rueApproximateBayesianInference2009}.
Values of $n_{t,\tau}$ are estimated using sampling. The total number of cases at week $t$ is then given by 
\begin{equation*}
n_{t} = \sum_{\tau} n_{t,\tau}.
\label{eqn:nttot}
\end{equation*}

We use the first twenty weeks of data in 2012 for training only, and begin generating estimates in epidemiological week 21 in 2012, which began on Sunday 20th May 2012. The model is fit to the data again every week, using all data available from the start of 2012 until week $t$. For efficiency, in fitting the model we discard all cases for which entry of the case into the surveillance system was delayed for over 26 weeks (i.e., half a year). We then set the maximum value of $\tau$ -- the number of weeks for which system entry was delayed -- to the number of weeks delay required to include 95\% of the remaining cases in training, or 8 weeks if this is greater. Remaining cases with a longer delay are omitted from training. The same approach is used for all of the following models, apart from the naive model.

\item[Google (Dengue).] This model is the same as the baseline model, with data on \emph{Google} searches related to the topic of \emph{dengue} added as an external regressor.
The mean $\lambda_{t,\tau}$ is now calculated as
\begin{equation*}
\log{(\lambda_{t,\tau})} = \mu + \alpha_t + \beta_{\tau} + \gamma^d\log{(G^{d}_t)},
\label{eqn:lambdattau.dengue}
\end{equation*}
where $G^{d}_t$ is the volume of \emph{Google} searches related to \emph{dengue} in week $t$ and $\gamma^d$ is a regression coefficient.

\item[Twitter.] This model is the same as the baseline model, with data on the volume of tweets that express personal experience of dengue added as an external regressor. The mean $\lambda_{t,\tau}$ is now calculated as
\begin{equation*}
\log{(\lambda_{t,\tau})} = \mu + \alpha_t + \beta_{\tau} + \delta\log{(T_t)},
\label{eqn:lambdattau.twitter}
\end{equation*}
where $T_t$ is the volume of \emph{Twitter} posts in week $t$ and $\delta$ is a regression coefficient.

\item[Google (Dengue) + Twitter.] This model is the same as the baseline model, with data on \emph{Google} searches related to the topic of \emph{dengue} and the volume of tweets that express personal experience of dengue added as external regressors.
The mean $\lambda_{t,\tau}$ is now calculated as
\begin{equation*}
\log{(\lambda_{t,\tau})} = \mu + \alpha_t + \beta_{\tau} + \gamma^d\log{(G^{d}_t)}. + \delta\log{(T_t)}.
\label{eqn:lambdattau.both}
\end{equation*}

\item[Google (all diseases).] This model is the same as the baseline model, with data on \emph{Google} searches related to the topics of \emph{dengue}, \emph{Zika} and \emph{chikungunya} added as external regressors.
The mean $\lambda_{t,\tau}$ is now calculated as
\begin{equation*}
\log{(\lambda_{t,\tau})} = \mu + \alpha_t + \beta_{\tau} + \gamma^d\log{(G^{d}_t)} + \gamma^z\log{(G^{z}_t)} + \gamma^c\log{(G^{c}_t)},
\label{eqn:lambdattau.alldiseases}
\end{equation*}
where $G^{z}_t$ and $G^{c}_t$ are the volumes of \emph{Google} searches in week $t$ related to \emph{Zika} and \emph{chikungunya}, and $\gamma^z$ and $\gamma^c$ are regression coefficients.

\item[Google (all diseases) + Twitter.] This model is the same as the baseline model, with data on \emph{Google} searches related to the topics of \emph{dengue}, \emph{Zika} and \emph{chikungunya} and the volume of tweets that express personal experience of dengue added as external regressors.
The mean $\lambda_{t,\tau}$ is now calculated as
\begin{equation*}
\log{(\lambda_{t,\tau})} = \mu + \alpha_t + \beta_{\tau} +
\gamma^d\log{(G^{d}_t)} + \gamma^z\log{(G^{z}_t)} + \gamma^c\log{(G^{c}_t)} + \delta\log{(T_t)}.
\label{eqn:lambdattau.alldiseasestw}
\end{equation*}

\item[Naive.] Following Yang et al. \cite{Yang2017}, this model uses the number of known cases for the previous week as the estimate of the number of dengue cases for the current week.
\end{description}

\subsection*{Evaluation of results}

We investigate two elements of model performance: accuracy and precision.

To evaluate accuracy, we consider the size of the \emph{prediction errors} generated by a model; that is, the difference between the number of dengue cases estimated by the model for a given week and the true number of cases in that week. A more accurate model would produce smaller prediction errors. To compare the size of prediction errors generated by different models, we calculate the \textit{mean absolute error} (MAE). This error metric is easy to interpret, as it is measured in numbers of dengue cases. 
In Fig. \ref{S1_Fig}, we discuss choice of error metric further and consider alternative metrics to the MAE.

To evaluate precision, we consider the size of the 95\% \emph{prediction intervals} generated by a model; that is, the size of the range of values within which the model estimates that there is a 95\% probability that the true number of dengue cases falls. A more precise model would generate smaller prediction intervals (assuming that 95\% of the true data points do fall within these prediction intervals, which we verify). To compare the size of prediction intervals generated by different models, we calculate the \textit{mean prediction interval} (MPI). We define the MPI as the mean width of the 95\% prediction interval for all estimates generated.

The dengue case count time series is characterised by a sequence of peaks and troughs. The error metrics we outline here will be affected by the model's performance during both peaks and troughs. However, accurate, precise information may be of most use to policymakers during epidemics when case counts are high. We therefore carry out sub-analyses in which we focus specifically on model accuracy and precision during periods of epidemics. To identify periods of epidemics, we apply the \textit{Moving Epidemic Method} (MEM \cite{Vega2013}) to historic data for Rio de Janeiro. This is a method which can be used to determine the minimum number of dengue cases per week that would be expected during epidemics. By applying this methodology to the official dengue case count data, we obtain an epidemic threshold of 550 dengue cases per week. Weekly counts below this threshold are considered inter-epidemic activity.

By adding online data streams to the models we consider, we introduce extra parameters into the models, potentially increasing the danger of overfitting. In our main error metric analyses, we test our models out-of-sample, thereby guarding against such overfitting as well as mimicking operational implementation. However, when evaluating the models, we also consider a further metric of model quality, the \textit{Watanabe-Akaike information criterion} (WAIC). This model quality metric rewards goodness of fit but explicitly penalises models for the presence of additional parameters. 

\section*{Results}

Following Yang et al. \cite{Yang2017}, we begin by comparing the accuracy of all models proposed to the accuracy of the naive model. Again, the naive model uses the known case count for the previous week as the estimate for the case count in the current week. To evaluate model accuracy, we calculate the \textit{mean absolute error} (MAE) for each model. To facilitate comparison of the models, we also calculate the \textit{relative MAE} for each model \cite{Reich2016a}. We define the relative MAE as the MAE of a given model divided by the MAE of the naive model. The relative MAE of the naive model is therefore 1. 

Table \ref{tab:maes.naive} shows that the naive model is vastly outperformed by all other models. The MAE for all other models is at least 37\% smaller than the MAE of the naive model. The best performing model is the \textit{Google (Dengue) and Twitter} model, for which the relative MAE is 0.502. As the performance of the naive model is considerably worse than all other models, we disregard it for further analyses.

\begin{table}[!ht]
\centering
\caption{{\bf Accuracy of all dengue nowcasting models compared to a naive model.} Following Yang et al. \cite{Yang2017}, we compare the accuracy of the naive model to all other models. We define the relative mean absolute error (relative MAE) as the MAE of a given model divided by the MAE of the naive model. The relative MAE of the naive model is therefore 1. We find that the naive model is vastly outperformed by all other models. Note that the baseline model is a more advanced model than the naive model, and is explicitly designed to account for the incremental delivery of the dengue case count data \cite{Bastos}. All models other than the naive model build on the baseline model. The best performing model is the \textit{Google (Dengue) and Twitter} model (bold), which exhibits an MAE 49.8\% smaller than that of the naive model.} 
\begin{tabular}{lcc}
\toprule
Model & MAE & relative MAE \\
\midrule
Baseline & 267.2 & 0.629 \\
\emph{Google (Dengue)} & 215.4 & 0.507\\
\emph{Twitter} & 223.3 & 0.525 \\
\textbf{\emph{Google (Dengue)} + \emph{Twitter}} & \textbf{213.3} & \textbf{0.502} \\
\emph{Google (all diseases)} & 218.8 & 0.515\\
\emph{Google (all diseases)} + \emph{Twitter} & 213.7 & 0.503\\
Naive & 425.0 & 1 \\
\bottomrule
\end{tabular}
\label{tab:maes.naive}
\end{table}

For the remainder of our analyses, we focus on comparing the models that use \textit{Google} and \textit{Twitter} data to the baseline model.  We redefine the relative MAE as the MAE of a given model divided by the MAE of the baseline model. The relative MAE of the baseline model is therefore 1. 

\begin{figure}[ht!]
\vspace{-3.5cm}
\thisfloatpagestyle{empty}
\centering 
\includegraphics[width = \linewidth]{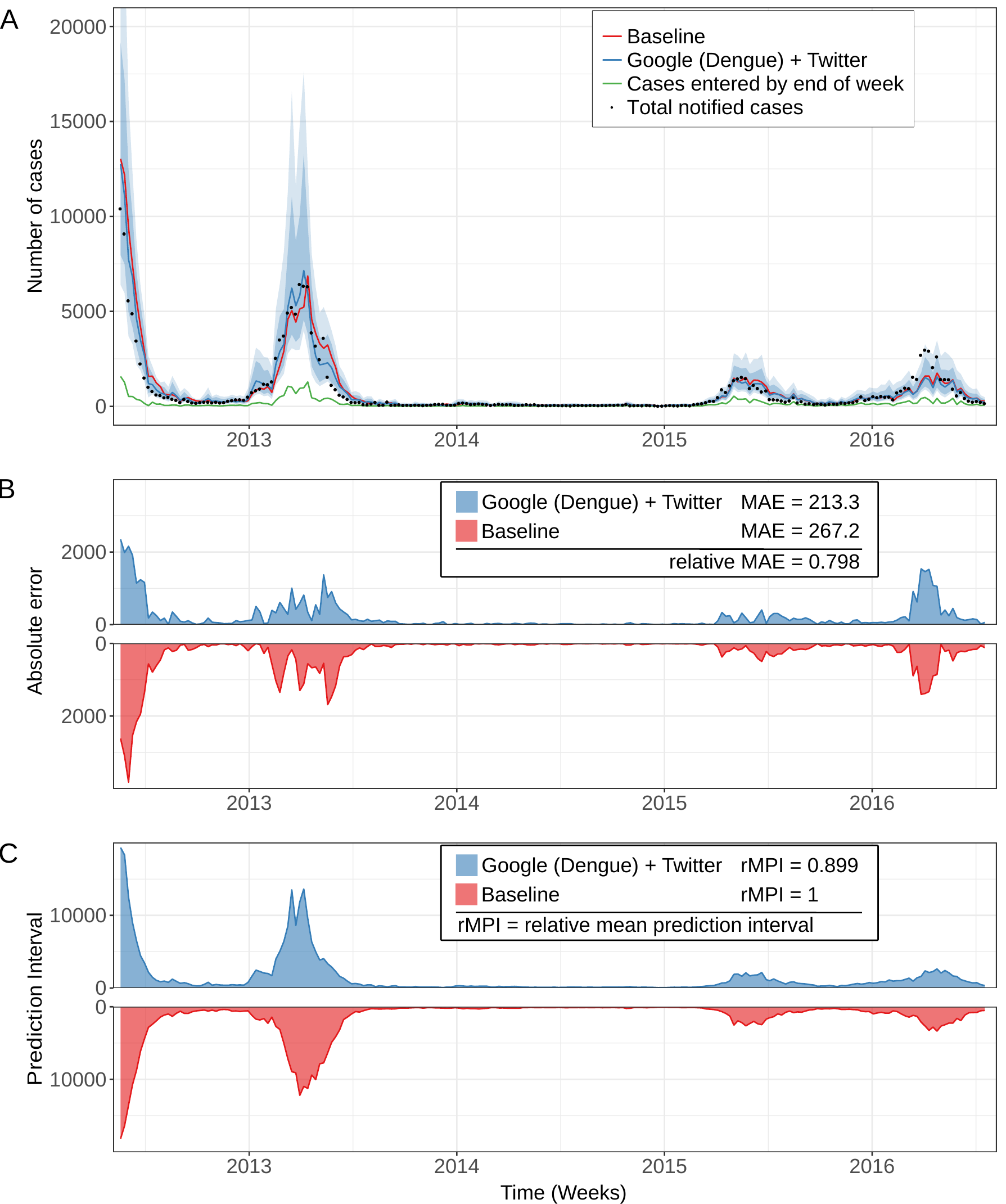}
\caption{{\bf Improving accuracy and reducing uncertainty for dengue case count estimates with \textit{Google} and \textit{Twitter}.} 
(A) We compare the performance of the baseline model with a model drawing on data from \textit{Google} and \textit{Twitter}. In black, we depict official data on the total number of dengue cases recorded for each week in Rio de Janeiro, from January 2012 until July 2016. In green, we depict the total number of dengue cases known to the authorities by the end of each week, which constitute a very small fraction of the total cases. In red, we depict estimates of the number of dengue cases generated by the baseline model for each week at the end of the corresponding week. The baseline model uses the official dengue case count data only and was designed to explicitly take into account the nature of the delays in the dengue data \cite{Bastos}, going beyond standard autoregressive approaches. It is clear that this model generally succeeds in capturing the timing and magnitude of the peaks. In blue, we depict estimates of the number of dengue cases generated by the \textit{Google (Dengue) + Twitter} model. It can be seen that the estimates enriched with \textit{Google} and \textit{Twitter} are often even closer to the final weekly dengue case count, in particular during the large peaks in case counts in 2012 and 2013. The blue shaded areas represent the 80\% (dark blue) and 95\% (light blue) prediction intervals for the \textit{Google (Dengue) + Twitter} model. (B) We compare the weekly absolute error for the baseline model and the \textit{Google (Dengue) + Twitter} model. While the mean absolute error (MAE) for the baseline model is 267.2 dengue cases per week, the MAE for the \textit{Google (Dengue) + Twitter} model is lower, at 213.3 dengue cases per week. The \textit{Google (Dengue) + Twitter} model is therefore more accurate. (C) An ideal model for estimating dengue case counts would produce accurate estimates with low uncertainty. To evaluate the level of uncertainty in the estimates produced by each model, we examine the \textit{relative mean prediction interval} (rMPI) for each model. We define the \textit{mean prediction interval} (MPI) as the mean width of the 95\% prediction interval for the full period for which estimates are generated. We define the rMPI as the MPI for the model divided by the MPI for the baseline model. The rMPI for the baseline model is therefore 1, whereas the rMPI for the \textit{Google (Dengue) + Twitter} model is lower at 0.899. The \textit{Google (Dengue) + Twitter} model therefore also generates more precise estimates.}
\label{fig:Fig3}
\end{figure}

Table \ref{tab:maes} shows that all the models enhanced with online data from either \textit{Google} or \textit{Twitter} outperform the baseline model. Across the full time period analysed, the baseline model exhibits an MAE of 267.2 cases. The model enhanced with data on tweets relating to dengue exhibits an MAE 16.4\% smaller than the baseline model, at 223.3 cases. The model enhanced with data on \textit{Google} searches relating to dengue exhibits an MAE 19.4\% smaller than the baseline model, at 215.4 cases. As was already seen in Table \ref{tab:maes.naive} however, the best performing model is the \emph{Google (Dengue) + Twitter} model, which draws on data on both \textit{Google} searches and tweets relating to dengue in tandem. This model exhibits an MAE of 213.3 cases, 20.2\% smaller than that of the baseline model (Fig. \ref{fig:Fig3}B). 
\begin{table}[!ht]
\centering
\caption{{\bf Accuracy of dengue nowcasting models using \textit{Google} and \textit{Twitter} data compared to the baseline model.} We redefine the relative mean absolute error (relative MAE) as the MAE of a given model divided by the MAE of the baseline model. The relative MAE of the baseline model is therefore 1. We find that all the models using online data outperform the baseline model. The best performing model is the \emph{Google (Dengue) + Twitter} model (bold), which exhibits an MAE 20.2\% smaller than that of the baseline model.}
\begin{tabular}{lcc}
\toprule
Model & MAE & relative MAE \\
\midrule
Baseline & 267.2 & 1 \\
\emph{Google (Dengue)} & 215.4 & 0.806 \\
\emph{Twitter} & 223.3 & 0.836 \\
\textbf{\emph{Google (Dengue) + Twitter}} & \textbf{213.3} & \textbf{0.798} \\
\emph{Google (all diseases)}& 218.8 & 0.819 \\
\emph{Google (all diseases)} + \emph{Twitter} & 213.7 & 0.800 \\
\bottomrule
\end{tabular}
\label{tab:maes}
\end{table}

The accuracy of estimates generated by the models which additionally draw on data on \textit{Google} searches relating to Zika and chikungunya is similar, with the \emph{Google (all diseases) + Twitter} model exhibiting an MAE of 213.7 cases, 20.0\% smaller than that of the baseline model. Overall, it therefore does not appear that integrating this extra \textit{Google} data relating to other arboviruses present in Rio de Janeiro improves accuracy of estimates of dengue incidence.

\begin{figure}[ht!]
\thisfloatpagestyle{empty}
\centering 
\includegraphics[width = \linewidth]{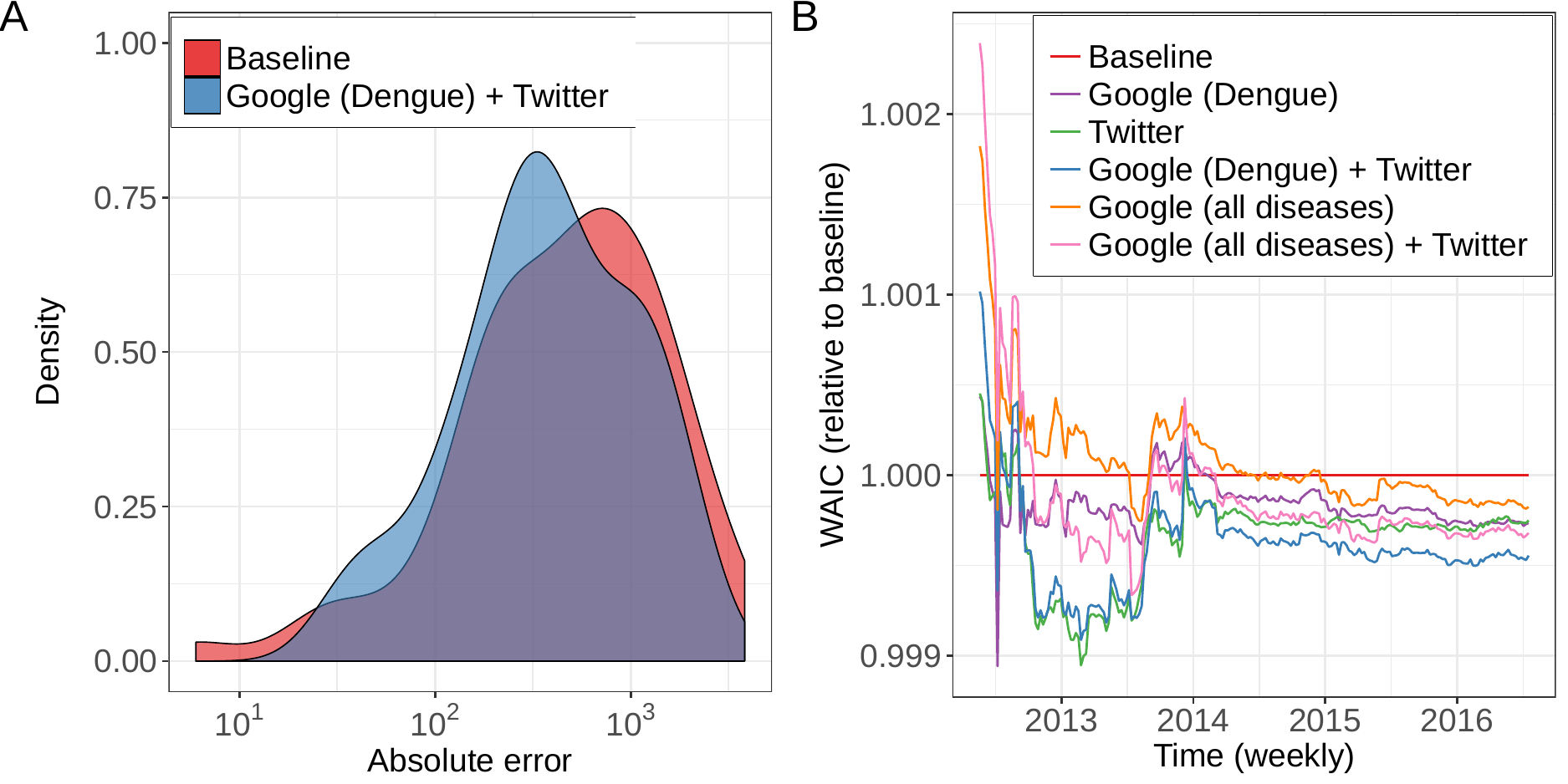}
\caption{{\bf Further analyses of the quality of dengue nowcasting models including \textit{Google} and \textit{Twitter} data.} (A) The performance of the models during epidemics is of particular importance. Using the \textit{Moving Epidemic Method} (MEM \cite{Vega2013}), we determine the epidemic threshold for Rio de Janeiro to be 550 dengue cases per week. For each week in which the final number of notified dengue cases was 550 or over, we determine the absolute error of the estimates generated by the baseline model and the \textit{Google (Dengue) + Twitter} model, and plot the distribution using a kernel density estimate. We find that the \textit{mean absolute error} (MAE) for the \textit{Google (Dengue) + Twitter} model (596.0 dengue cases per week; blue) is again considerably lower than the MAE for the baseline model (774.8 dengue cases per week; red). (B) In addition to evaluating the accuracy and precision of out-of-sample estimates generated by the models, here we examine a further metric of model quality, the \textit{Watanabe-Akaike information criterion} (WAIC). The WAIC rewards goodness of fit but explicitly penalises models for the presence of additional parameters, such as data on \textit{Google} searches or tweets. We evaluate the quality of all six models explored in our main analysis: the baseline model (red), the \textit{Google (Dengue)} model (purple), the \textit{Twitter} model (green), the \textit{Google (Dengue) + Twitter} model (blue), the \textit{Google (all diseases)} model (orange) and the \textit{Google (all diseases) + Twitter} model (pink). As the model is fit each week when new data arrives, we calculate a WAIC value for each of the six models for every week. To facilitate comparison of these weekly WAIC values, for each week we normalise the six WAIC values by the WAIC for the baseline model. The resulting value for the baseline model is therefore always 1 (red line). A lower WAIC indicates a higher quality model. It can be observed that the models enhanced by online data generally exhibit lower WAIC values than the baseline model. We note that, again, the \textit{Google (Dengue) + Twitter} model (blue) performs particularly well.}
\label{fig:Fig4}
\end{figure}

The performance of the models during epidemics is of particular importance. We therefore examine whether the estimates generated by the \textit{Google (Dengue) + Twitter} model are more accurate when considering periods of epidemics alone. Using the \textit{Moving Epidemic Method} (MEM \cite{Vega2013}), we determine the epidemic threshold for Rio de Janeiro to be 550 dengue cases per week. For each week in which the final number of notified dengue cases was 550 or over, we calculate the absolute error of the estimates generated by the baseline model and the \textit{Google (Dengue) + Twitter} model. We find that during epidemics, the baseline model exhibits an MAE of 774.8 cases. In contrast, the \textit{Google (Dengue) + Twitter} model exhibits an MAE of 596.0 cases, 23.1\% lower than the baseline model (Fig. \ref{fig:Fig4}A).

\FloatBarrier

The inclusion of extra parameters in a model, such as data on \textit{Google} searches or tweets, increases the likelihood of overfitting. While the analyses detailed so far have considered estimates generated out-of-sample, thereby guarding against this danger, we also calculate the \textit{Watanabe-Akaike information criterion} (WAIC) model quality metric for each of our six models. The WAIC rewards goodness of fit whilst penalising models for the inclusion of extra parameters. As the model is fit each week when new data arrives, we calculate a WAIC value for each of the six models for every week. 

Fig. \ref{fig:Fig4}B depicts the weekly WAIC values for all six models, relative to the baseline model. 
A lower WAIC value indicates a higher quality model. We find that models enhanced by online data generally exhibit lower WAIC values than the baseline model. In most weeks, the lowest WAIC is again obtained by the \emph{Google (Dengue) + Twitter} model, which draws on data on both \textit{Google} searches and tweets relating to dengue in tandem.

An ideal model for estimating current dengue case counts would not only produce accurate estimates, but would also produce precise estimates, where uncertainty about the true value was low. We therefore examine whether dengue nowcasting models enhanced by online data generate estimates that are more precise, as well as more accurate. To evaluate the precision of estimates produced by each model, we calculate the \textit{mean prediction interval} (MPI), the mean width of the 95\% prediction interval for all estimates generated. To facilitate comparison to the performance of the baseline model, we also calculate the \textit{relative MPI} (rMPI), which we define as the MPI for a given model divided by the MPI for the baseline model. The rMPI for the baseline model is therefore 1. 

Table \ref{tab:shrinkings} shows that the rMPI for all models enhanced by online data is lower than 1. This indicates that the estimates generated by the models enhanced by online data are more precise than those generated by the baseline model. The \textit{Twitter} model is the most precise model, exhibiting an MPI which is 11.1\% lower than the MPI of the baseline model. The \textit{Google (Dengue)} model, drawing on data on \textit{Google} searches relating to dengue, achieves a smaller but still notable improvement of 8.8\%. The combined \textit{Google (Dengue) + Twitter} model, which produced the most accurate estimates, generates the second most precise estimates, with an MPI 10.1\% lower than the the MPI of the baseline model (Fig. \ref{fig:Fig3}C). 

\begin{table}[!ht]
\caption{{\bf Precision of dengue nowcasting models using \textit{Google} and \textit{Twitter} data compared to the baseline model.} We define the mean prediction interval (MPI) as the mean width of the 95\% prediction interval for all estimates generated. The MPI for the baseline model is given in parentheses. We define the relative mean prediction interval (rMPI) as the MPI for the model divided by the MPI for the baseline model. The rMPI for the baseline model is therefore 1. We find that models using online data generate more precise estimates, reflected by lower rMPIs. The most precise model is the \textit{Twitter} model (bold), followed by the \textit{Google (Dengue) + Twitter} model. We also verify that the 95\% prediction intervals reliably represent the range within which 95\% of true data points fall. We find that whether considering all weeks, weeks with more than 550 cases (i.e., during epidemics) or weeks with fewer than 550 cases (i.e., outside epidemics), the 95\% prediction intervals appear to behave as desired.}
\centering
\begin{tabular}{lcccc}
\toprule
& relative Mean &\multicolumn{3}{c}{Percentage points within} \\
Model & Prediction Interval &\multicolumn{3}{c}{95\% prediction interval } \\
\cmidrule{3-5}\
&& all & $>550$ & $<550$ \\
\midrule
Baseline 				& 1 (1554.6) & 95.0 & 93.7 & 95.5 \\
\emph{Google (Dengue)} 	& 0.912 & 94.5 & 93.7 & 94.8 \\
\emph{Twitter} 		& \textbf{0.889} & 95.4 & 96.9 & 94.8 \\
\emph{Google (Dengue)} + \emph{Twitter} & 0.899 & 94.5 & 95.3 & 94.2 \\
\emph{Google (all diseases)} & 0.938 & 95.4 & 96.9 & 94.8\\
\emph{Google (all diseases)} + \emph{Twitter} & 0.901 & 95.4 & 95.3 & 95.5 \\
\bottomrule
\end{tabular}
\label{tab:shrinkings} 
\end{table}

The precision of estimates generated by models which additionally draw on data on \textit{Google} searches relating to Zika and chikungunya is again similar, with the \emph{Google (all diseases) + Twitter} model exhibiting an MPI 9.9\% lower than the the MPI of the baseline model. It therefore does not appear that integrating this extra \textit{Google} data relating to other arboviruses present in Rio de Janeiro improves the precision of estimates of dengue incidence.

We verify whether the 95\% prediction intervals continue to reliably represent the range within which 95\% of true data points fall. Table \ref{tab:shrinkings} demonstrates that whether considering all weeks, weeks with more than 550 cases (i.e., during epidemics) or weeks with fewer than 550 cases (i.e., outside epidemics), the 95\% prediction intervals appear to behave as desired. In other words, this 10\% improvement in the precision of estimates does not come at the cost of the reliability of the prediction intervals.

The characteristics of the dengue season in Rio de Janeiro vary from year to year. In some years, over 5\,000 cases a week are reported at the height of the season, whereas in other years, the case count is much lower (Fig. \ref{fig:Fig1}). In addition, previous research has highlighted that the relationship between online data and case counts may vary across time \cite{Preis2014}. We therefore investigate whether the use of online data helps deliver more accurate estimates of dengue incidence in Rio de Janeiro in each of the years covered in our analysis. 

In Table \ref{tab:maes.years}, we report the relative MAE for each model for each year of analysis. We note that statistics for 2012 and 2016 are based on incomplete years, as the analyses begin in Week 21 of 2012 and end in Week 29 of 2016. We find that in 2012, 2013, 2014 and 2015, the accuracy of all models using online data is greater than the accuracy of the baseline model. Using the \emph{Google (Dengue) + Twitter} model, the MAE is reduced by between 11\% and 32\%.

\begin{table}[!ht]
\caption{{\bf Evaluating the accuracy of dengue nowcasting models using \textit{Google} and \textit{Twitter} across different years.} For each year, we define the relative mean absolute error (relative MAE) as the MAE of a given model divided by the MAE of the baseline model. The MAE is given in parentheses. In bold, we highlight the lowest relative MAE for each year. We find that in 2012, 2013, 2014 and 2015, the accuracy of all models using online data is greater than the accuracy of the baseline model. In 2016, we find that the baseline model delivers the most accurate estimates. However, Fig. \ref{fig:Fig3}A shows that in 2016, the performance of the baseline model itself is notably worse than in previous years. We discuss the particular circumstances of 2016 in more detail in the text.}
\begin{adjustwidth}{-2.25in}{0in}
\flushright
\begin{tabular}{llllll}
\toprule
&\multicolumn{5}{c}{Relative mean absolute error}\\
\cmidrule{2-6}
Model & 2012 & 2013 & 2014 & 2015 & 2016 \\
\midrule
Baseline 										& 1 (678.4) 	& 1 (354.3) 	& 1 (19.0) 		& 1 (123.0)		& \textbf{1 (369.1)} \\
\emph{Google (Dengue)} 							& 0.69 			& 0.76 			& 0.98 			& 0.93 			& 1.03 \\
\emph{Twitter} 									& 0.74 			& 0.78 			& 0.91 			& 0.96 			& 1.03 \\
\emph{Google (Dengue)} + \emph{Twitter} 		& 0.68 			& \textbf{0.74} & \textbf{0.86} & \textbf{0.89} & 1.08 \\
\emph{Google (all diseases)}					& 0.73 			& 0.77 			& 0.96 			& 0.90 			& 1.02 \\
\emph{Google (all diseases)} + \emph{Twitter} 	& \textbf{0.65} & 0.80 			& 0.87 			& 0.92 			& 1.02 \\
\bottomrule
\end{tabular}
\end{adjustwidth}
\label{tab:maes.years}
\end{table}

In 2016 however, we find that the baseline model delivers the most accurate estimates, and that the MAE of estimates generated by the \emph{Google (Dengue) + Twitter} model is 8\% higher. At the same time however, we note that the MAE for the baseline model in 2016 (369.1 cases per week) is relatively high given the size of the peak. For example, the MAE for the baseline model in 2013 was similar at 354.3 cases per week, but the peak number of dengue cases per week in 2013 was 6430 in comparison to a peak of 2973 cases per week in 2016. This diminished performance in 2016 can also be seen in Fig. \ref{fig:Fig3}A.

Why might we observe differing results for 2016 in comparison to earlier years? A potential answer to this question can be found by examining the nature of the delays in the entry of dengue cases into the surveillance system around this period. Figure~\ref{S3_Fig} illustrates that from January 2012 to May 2015, there was a mean delay of 4.9 weeks until 80\% of dengue cases for a given week were entered into the surveillance system, with a standard deviation of 1.5 weeks. From June 2015 to December 2015 however, delays were notably reduced such that there was a mean delay of 2 weeks until 80\% of dengue cases for a given week were entered into the surveillance system. 
From January 2016 to the end of the dataset in July 2016, the delays increased again to a mean of 4.6 weeks until 80\% of dengue cases for a given week were entered into the surveillance system. 
This abnormally large variation in delays may have made it particularly difficult for the baseline model to correctly model the delay structure, leading to a higher baseline MAE for 2016.

It is also worth noting that there was a Zika outbreak in Brazil during the 2016 dengue season. Zika is not only spread by the same mosquito as dengue, but also shares some symptoms. Difficulty in discerning the symptoms of dengue from the symptoms of Zika before a laboratory analysis has taken place will have led to some cases of dengue being recorded as suspected cases of Zika, and vice versa. The Zika outbreak was also covered widely in the media, and it is possible that people with dengue may have searched for information relating to Zika instead. Fig. \ref{fig:Fig1}D shows that there was a surge in searches relating to Zika in 2016, and Fig. \ref{fig:Fig1}E shows that a similar surge occurred for searches relating to a further arbovirus present in Rio de Janeiro, chikungunya. Indeed, Table \ref{tab:maes.years} shows that for 2016, the best performing models using online data are the \emph{Google (all diseases)} model and the \emph{Google (all diseases) + Twitter} model, both of which additionally draw on data on \textit{Google} searches relating to Zika and chikungunya. However, both models still generate estimates with errors which were 2\% greater than the errors generated by the baseline model.

\section*{Discussion}
Here, we investigate whether data on \emph{Google} searches and \emph{Twitter} posts relating to dengue can be used to improve nowcasts of dengue case counts, when official case count data is not only delayed but also released incrementally, as is frequently the case. Using Rio de Janeiro in Brazil as a case study, we present analyses which show that by drawing on \emph{Google} and \emph{Twitter} data in parallel, weekly estimates of the current number of dengue case counts can be made both more accurate and more precise than estimates that use historic official data alone. The explicit modelling of the true incremental delivery of the case count data means that this approach can be used in practice, with no need to aggregate data up to a coarser temporal granularity such as months. Our results also illustrate the potential value of considering multiple online data streams in parallel, instead of focusing on the relationship between case count data and one online data stream alone.

The only year in which we find that online data does not improve estimates is 2016, when there was also a Zika outbreak in Rio de Janeiro. As Zika and dengue share symptoms, it is possible that people were searching for information about one disease when they were suffering from the other. Future work could look to build a combined model of the incidence of the arboviruses dengue, Zika and chikungunya, to better exploit the relationships between the three diseases that exist in both case count and online data. An extended model could also look to draw on other rapidly available data sources, such as weather data \cite{Luz2008,Hii2012,Ramadona2016}. The framework described here has been developed for use in the \emph{InfoDengue} surveillance system, used in hundreds of Brazilian cities \cite{codeco2016}. Extensions of this work could also verify whether this online data approach would benefit other cities and countries too.

Dengue is a global burden, and a lack of timely data on case counts leaves policymakers without the information they need to intervene early in an outbreak. We hope that careful development of analysis frameworks to exploit rapidly available alternative data sources, integrated into surveillance systems such as \emph{InfoDengue}, will help mitigate this problem.

\section*{Acknowledgements}
{\small
GM acknowledges EPSRC grant EP/L015374/1. TP and HSM were supported by Research Councils UK grant EP/K039830/1, the University of Warwick Brazil Partnership Fund, and The Alan Turing Institute under the EPSRC grant EP/N510129/1 (awards TU/B/000006 and TU/B/000008). GM, TP and HSM are also grateful for support provided by the University of Warwick GRP Behavioural Science. LSB acknowledges CAPES grant 88881.068124/2014-01 and FAPERJ E-26/201.277/2021. CTC acknowledges CNPq grant 305553/2014-3 and InfoDengue support from the SVS/Brazilian Ministry of Health. The authors are grateful to the Secretaria Municipal de Sa\'{u}de do Rio de Janeiro for providing access to the data on dengue cases and to the Observat\'{o}rio da Dengue (UFMG) for the data on the volume of tweets related to dengue.}

\clearpage

\makeatletter
\setlength{\@fptop}{0pt}
\makeatother

\FloatBarrier
\section*{Supplementary information}

\setcounter{figure}{0}
\renewcommand{\thefigure}{S\arabic{figure}}

\noindent
Dengue case count time series are characterised by a sequence of peaks and troughs. The vast differences in case counts at different points in the time series can pose challenges for the evaluation of models that seek to estimate these case counts\cite{Hyndman2006,Reich2016a}.

In our main analysis, we use the \textit{mean absolute error (MAE)} metric to evaluate the performance of our model. This error metric is easy to interpret, as it is measured in numbers of dengue cases. For example, across the full time period analysed in this paper, the baseline model exhibits a mean absolute error of 267.2 dengue cases per week. The mean absolute error also gives equal weight to underestimates and overestimates of the same size; in other words, it is \textit{symmetric}. This is an advantage in comparison to other common error metrics such as the \textit{mean absolute percentage error (MAPE)}. The mathematical properties of the MAPE allow it to vary from $-100\%$ to $+\infty\%$. For the same reasons, the MAPE gives less weight to underestimates than overestimates.

However, in evaluating the performance of a model, it might be desirable to consider whether an error of a given number of cases occurred when the true number of cases was very high or very low. The mean absolute error does not behave like this, and allocates an error of a given number of cases the same weight at a peak and at a trough. For this reason, we also consider an alternative error metric, the \textit{logarithmic error} \cite{Tofallis2015}. Unlike the mean absolute error, the logarithmic error is not scale-dependent: that is, it is not defined in the units of the underlying time series, and the metric takes into account the size of the corresponding true value. Like the mean absolute error however, the logarithmic error is symmetric.

\begin{figure}[!h]
\thisfloatpagestyle{empty}
\centering \includegraphics[width = \linewidth]{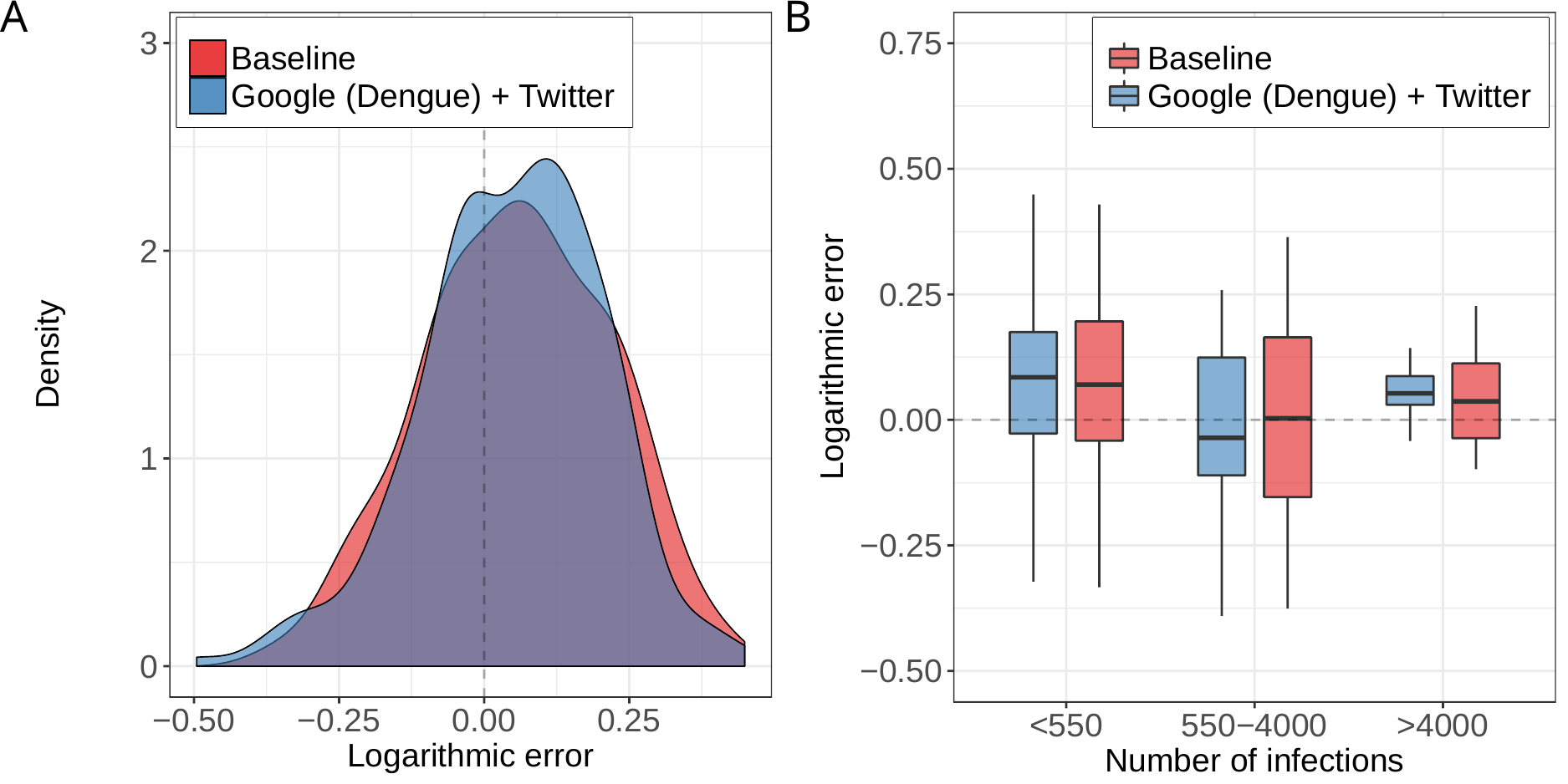}
\caption{
{\bf Accuracy of the baseline and \textit{Google (Dengue) + Twitter} models according to the logarithmic error metric.}
(A) In evaluating the accuracy of our models, we also consider an alternative error metric, the \textit{logarithmic error}. Unlike the \textit{mean absolute error} metric that we use in our main analysis, the logarithmic error metric takes into account whether an error of a given number of cases occurred when the true number of cases was very high or very low. The logarithmic error is defined as $\log_{10}{Q}$ where $Q=\hat{y}/y$, $\hat{y}$ is the predicted value and $y$ the true value. Both the baseline and the \textit{Google (Dengue) + Twitter} models show a clear tendency to overestimate rather than underestimate the case counts. Errors generated by the \textit{Google (Dengue) + Twitter} model are slightly lower overall, and therefore more concentrated around $0$. The error distribution is plotted using a kernel density estimate.
(B) We evaluate how the error distribution differs in periods of epidemics and outside such periods. Using the \textit{Moving Epidemic Method} (MEM \cite{Vega2013}), we determine the epidemic threshold for Rio de Janeiro to be 550 dengue cases per week. We also investigate how the error distribution changes when dengue case counts are particularly high. Here, we use a threshold of 4\,000 dengue cases per week. We find that below the epidemic threshold of 550 dengue cases per week, both models generally overestimate the number of dengue cases. Above the epidemic threshold, other than in periods when dengue case counts are particularly high, we find that both models tend to slightly underestimate the number of dengue cases. When case counts are higher than 4\,000 a week, the models tend to slightly overestimate the number of dengue cases again, but error rates are relatively low in the context of the true dengue case counts. In all three scenarios, errors generated by the \textit{Google (Dengue) + Twitter} model tend to be lower than errors produced by the baseline model.}
\label{S1_Fig}
\end{figure}

The logarithmic error is defined as $\log_{10}{Q}$ where $Q=\hat{y}/y$, $\hat{y}$ is the predicted value and $y$ the true value \cite{Tofallis2015}. Both the baseline and the \textit{Google (Dengue) + Twitter} models show a clear tendency to overestimate rather than underestimate the case counts (Fig. \ref{S1_Fig}A). Errors generated by the \emph{Google (Dengue) + Twitter} model are slightly lower overall, and therefore more concentrated around $0$.

We further evaluate how the error distribution differs in periods of epidemics and outside such periods. To identify periods of epidemics, we apply the \textit{Moving Epidemic Method} (MEM \cite{Vega2013}) to historic data for Rio de Janeiro. This is a method which can be used to determine the minimum number of dengue cases per week that would be expected during epidemics. Identifying an `epidemic threshold' is of use when interpreting incoming disease surveillance data, to determine whether an increase in case counts is likely to simply reflect fluctuations in baseline disease incidence, or might indicate the onset of an epidemic. The full methodology is detailed in the MEM paper \cite{Vega2013}. By applying this methodology to the dengue data for Rio de Janeiro, we obtain an epidemic threshold of 550 dengue cases per week. We also investigate how the error distribution changes when dengue case counts are particularly high. Here, we use a threshold of 4\,000 dengue cases per week.

Figure \ref{S1_Fig}B shows how the error distributions vary for the two models in these three periods: periods outside epidemics, when weekly case counts are below 550; and two classes of periods during epidemics, firstly when weekly case counts are below 4\,000, and secondly when weekly case counts are particularly high and above 4\,000. We find that below the epidemic threshold of 550 dengue cases per week, both models generally overestimate the number of dengue cases. Above the epidemic threshold, other than in periods when dengue case counts are particularly high, we find that both models tend to slightly underestimate the number of dengue cases. When case counts are higher than 4\,000 a week, the models tend to slightly overestimate the number of dengue cases again, but error rates are relatively low in the context of the true dengue case counts. In all three scenarios, errors generated by the \textit{Google (Dengue) + Twitter} model tend to be lower than errors produced by the baseline model.

\begin{figure}[!h]
\thisfloatpagestyle{empty}
\centering \includegraphics[width = \linewidth]{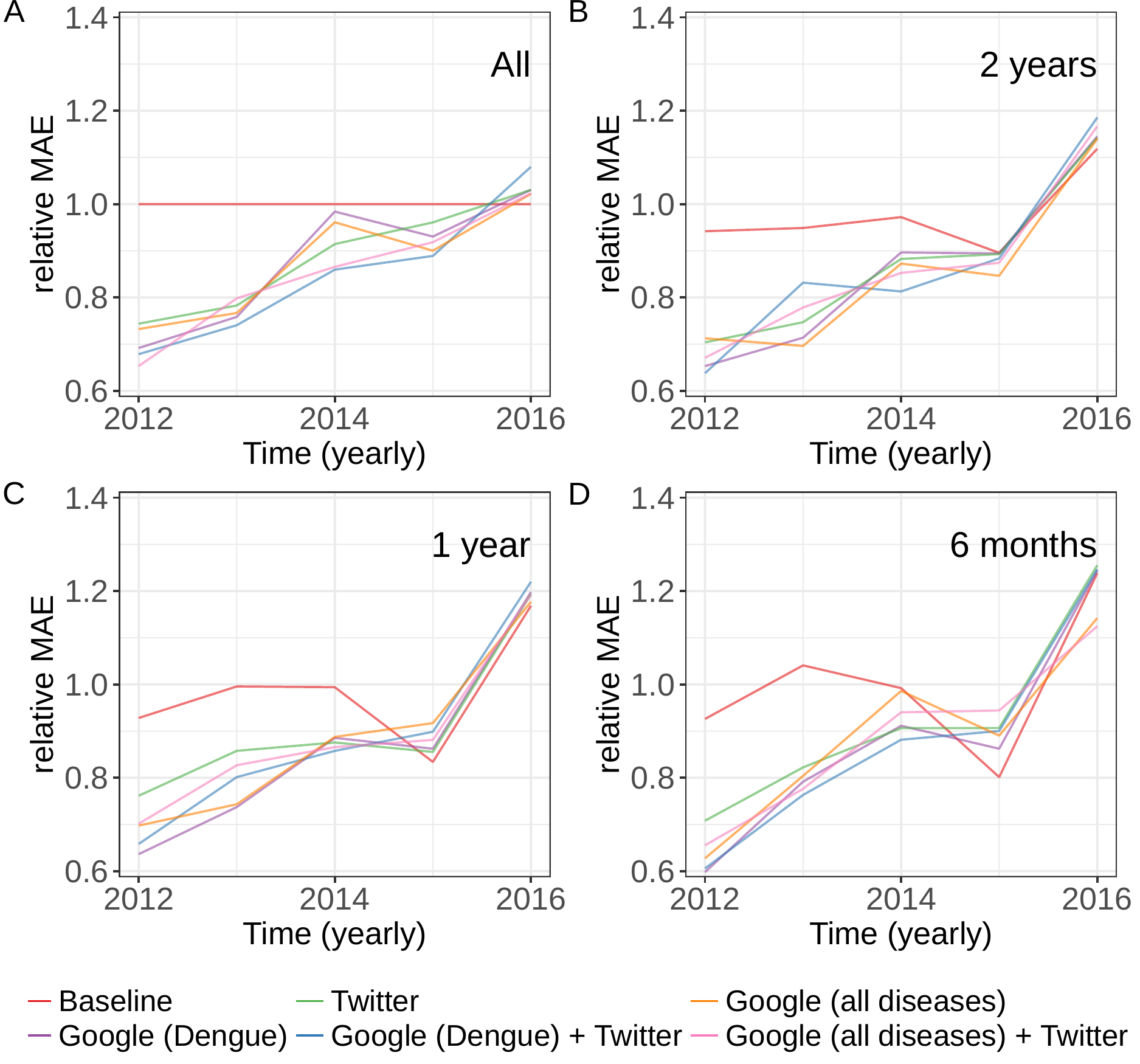}
\caption{
\textbf{Exploring the benefits of using a sliding training window.} The parameters for the models we have described so far are fit using all data available from the beginning of the time series in 2012 to the current week. This means that as each week passes, the volume of data on which the model is trained grows. Here, we explore whether there is any benefit to only considering recent data when fitting the model, building on a previous study into the relationship between online data and influenza incidence \cite{Preis2014}. We consider the performance of all six models explored in our main analysis: the baseline model (red), the \textit{Google (Dengue)} model (purple), the \textit{Twitter} model (green), the \textit{Google (Dengue) + Twitter} model (blue), the \textit{Google (all diseases)} model (orange) and the \textit{Google (all diseases) + Twitter} model (pink). To evaluate performance, we calculate the \textit{relative mean absolute error} (relative MAE), which we define as the mean absolute error (MAE) of a given model divided by the MAE of the baseline model when trained on all data available from the beginning of the time series to the current week. We assess the relative MAE of each model for each year from 2012 to 2016 (where data for 2016 is partial, ending in mid-July). (A) Yearly performance of all six models when trained on all data available from the beginning of the time series to the current week. As the baseline model is our reference model, the relative MAE for the baseline model is 1 for each year. (B) Yearly performance of all six models when using a sliding training window of two years. In other words, in each week, the model is trained on the most recent two years of data. (C) Yearly performance of all six models when using a sliding training window of one year. (D) Yearly performance of all six models when using a sliding training window of six months. Overall, we find little evidence of any consistent benefit of using only recent data to train the model.}
\label{S2_Fig}
\end{figure}

\pagebreak

\begin{figure}[!h]
\thisfloatpagestyle{empty}
\centering \includegraphics[width = 0.9\linewidth]{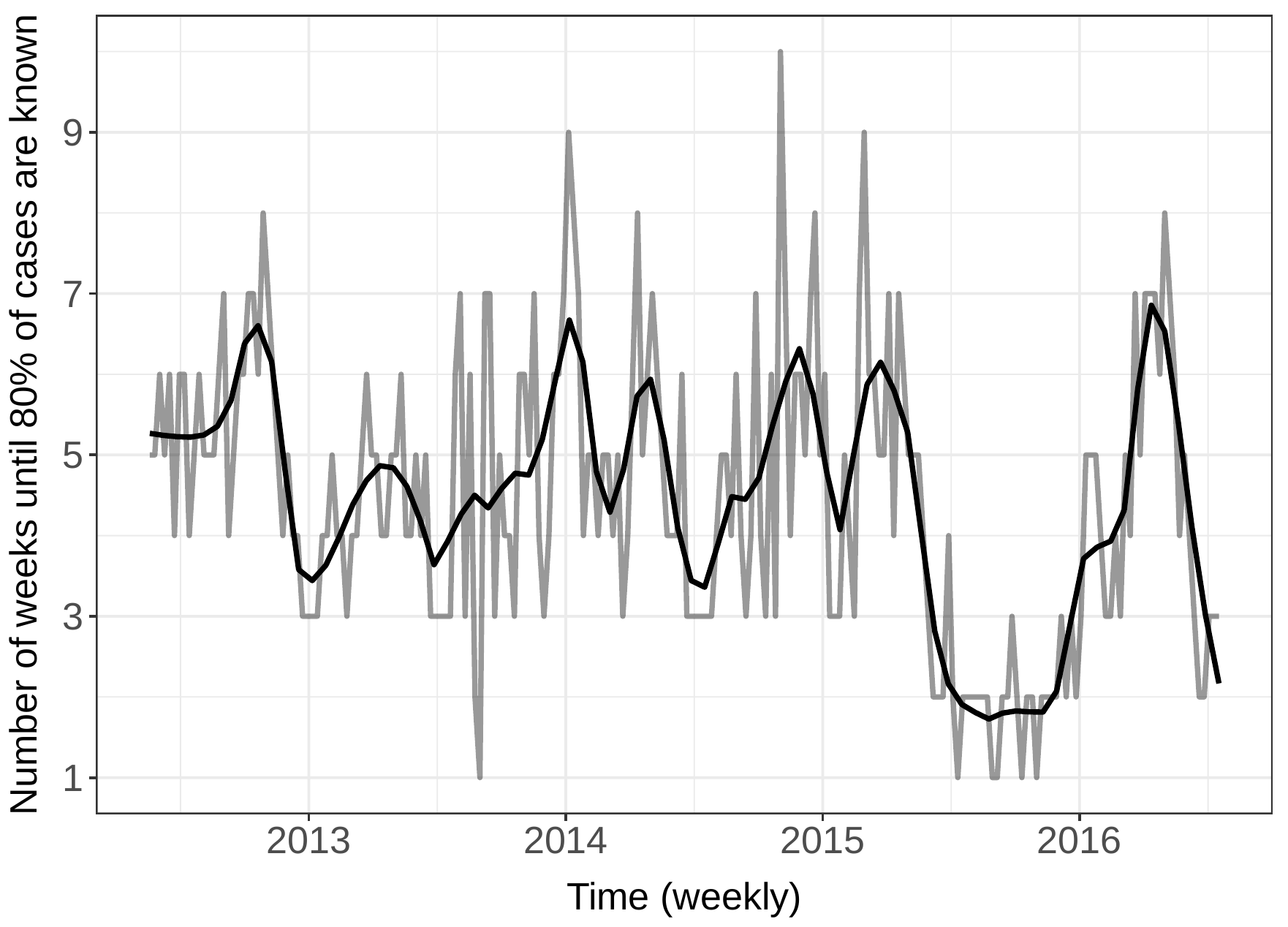}
\caption{
\textbf{Abnormal variation in delays in the recording of dengue cases before the 2016 season.} In 2016, we observe diminished performance of the baseline model in comparison to earlier years (Fig. 3A). To try and explain this finding, we investigate whether there were any changes in the nature of the delays in the entry of dengue cases into the surveillance system around this period. We find that from January 2012 to May 2015, there was a mean delay of 4.9 weeks until 80\% of dengue cases for a given week were entered into the surveillance system, with a standard deviation of 1.7 weeks. From June 2015 to December 2015 however, delays were notably reduced, such that there was a mean delay of 2 weeks until 80\% of dengue cases for a given week were entered into the surveillance system. From January 2016 to the end of the dataset in July 2016, the delays increased again to a mean of 4.6 weeks until 80\% of dengue cases for a given week were entered into the surveillance system. This abnormally large variation in delays may have made it particularly difficult for the baseline model to correctly model the delay structure, leading to worse performance in 2016.}
\label{S3_Fig}
\end{figure}

\end{document}